\documentclass[a4paper,11pt]{article}
\pdfoutput=1 

\usepackage{jcappub} 

\usepackage[T1]{fontenc} 
\usepackage{caption}

\newcommand{\beq}{\begin{equation}}
\newcommand{\eeq}{\end{equation}}
\newcommand{\bea}{\begin{eqnarray}}
\newcommand{\eea}{\end{eqnarray}}
\newcommand{\beqn}{\begin{equation*}}
\newcommand{\eeqn}{\end{equation*}}
\newcommand{\bean}{\begin{eqnarray*}}
\newcommand{\eean}{\end{eqnarray*}}

\newcommand*{\cref}[1]{Chapter~\ref{#1}}
\DeclareMathOperator{\sech}{sech}

\title{\boldmath Model independent results for the inflationary epoch and the breaking of the degeneracy of models of inflation}

\author[a]{Gabriel Germ\'an}

\affiliation[a]{Instituto de Ciencias F\'{\i}sicas, Universidad Nacional Aut\'onoma de M\'exico\\ Cuernavaca, Morelos, 62210, Mexico}

\emailAdd{gabriel@icf.unam.mx}

\abstract{We address the problem of determining inflationary characteristics in a model independent way. We start from a recently proposed equation which allows to accurately calculate the value of the inflaton at horizon crossing $\phi_k$. We then use an equivalent form of this equation to write a formula that relates the number of e-folds from horizon crossing to the pivot scale $N_{ke}+N_{ep}$ with the tensor-to-scalar index $r$, hence a general bound for $N_{ke}+N_{ep}$ follows. $N_{ke}$ is the number of e-folds from the scale factor $a_k$ during inflation to the end of inflation at $a_e$ and $N_{ep}$ is the number of e-folds from $a_e$ to the pivot scale factor  $a_p$.
In particular, at present $r < 0.063$ implies $N_{ke}+N_{ep}< 112.5$ e-folds at $k=k_p$ and 128.1 e-folds at the present scale with wavenumber mode $k_0$. We also give a lower bound to the size of the universe during the inflationary epoch that gave rise to the current observable universe. We also discussed the problem of degeneracy of inflationary models and argue that this degeneration can only be resolved by studying model predictions from the reheating epoch.}

\begin{document}
\maketitle
\flushbottom

\section {\bf Introduction}\label{Int}

During the last several years we have seen an extraordinary advance in our knowledge of the universe, its composition, geometry and evolution. The idea of an inflationary universe remains solid some 40 years after its inception \cite{Guth:1980zm}, \cite{Linde:1984ir}, (for reviews see e.g., \cite{Lyth:1998xn}, \cite{Baumann:2009ds}, \cite{Martin:2018ycu}), however the existence of a plethora of models \cite{Martin:2013tda} constantly reminds us that our knowledge of that epoch is imprecise, and even more so when we consider the time of reheating after inflation ends, for reviews on reheating see e.g., \cite{Bassett:2005xm},  \cite{Allahverdi:2010xz}, \cite{Amin:2014eta}. Numerous works have been done in our attempt to better understand the reheating era with varying degrees of success \cite{Liddle:2003as} - \cite{German:2020kdp} . In this work, we initially address the problem of determining important inflationary characteristics in a model independent way and then study how the degeneracy of inflationary models can possibly be resolved by considering reheating.

The organization of the article is as follows: in Section \ref {RI} we first start from a recently proposed equation  \cite{German:2020dih}  which allows us to accurately calculate the value of the inflaton at horizon crossing $\phi_k$. We then use an equivalent form of this equation to write a formula that relates the  number of e-folds $N_{ke}+N_{ep}$, from $a_k$ during inflation to the pivot scale at $a_p$, to the tensor-to-scalar ratio $ r $ hence a general bound for $N_{ke}+N_{ep}$ follows. $N_{ke}$ is the number of e-folds from the scale factor $a_k$ during inflation to the end of inflation at $a_e$ and $N_{ep}$ is the number of e-folds from $a_e$ to the pivot scale factor  $a_p$. In particular, for the present bound $r < 0.063$   \cite{Aghanim:2018eyx}, \cite{Akrami:2018odb} we get $N_{ke}+N_{ep} < 112.5$ e-folds at $k=k_p$ or 128.1 at $k=k_0$. We end the section by calculating a lower bound to the size of the universe, during the inflationary epoch, that gave rise to the current observable universe. 
In Section \ref {RR} we discuss the reheating epoch and give formulas for the number of e-folds during reheating and during the radiation dominated epochs.
In Section \ref {MUT} we study three models of inflation which are well approximated around the origin by a quadratic monomial and can be described by an equation of state (EoS) during reheating given by $\omega_{re} =0$. We discuss how these models are degenerated during the inflationary epoch and argue that the breaking of this degeneracy is only possible by the study of their predictions for the reheating epoch. Finally in Section \ref {Con} we give our conclusions on the most important points discussed in the article.
\section {\bf The total number of e-folds}\label{RI}
The equation which determines the inflaton field $\phi$ at horizon crossing follows from considering the number of e-folds that passes from the moment the scale with wavenumber mode $k_p\equiv a_p H_p$ exit the horizon during inflation until that same scale re-enters the horizon i.e., $\ln(\frac{a_p}{a_k})=N_{ke}+N_{ep}$ where $N_{ke}\equiv \ln\frac{a_e}{a_k}$ is the number of e-folds from $\phi_k$ up to the end of inflation at $\phi_e$  and $N_{ep}\equiv \ln\frac{a_p}{a_e}$ is the {\it postinflationary} number of e-folds from the end of inflation at $a_e$ up to the pivot scale factor $a_p$. In the equation above, multiplying $\frac{a_p}{a_k}$ above and below by $H_k$ and setting $k\equiv a_k H_k=k_p$ we get \cite{German:2020dih} 
\begin{equation}
\ln[\frac{a_{p}H_k}{k_p}]=N_{ke}+N_{ep}\;.
\label{EQ}
\end{equation}
The Hubble function at $k$ is given by $H_k=\sqrt{8\pi^2 \epsilon_k A_s}$, notice that the Hubble function  introduces the scalar power spectrum amplitude given here by $A_s$. Eq.~\eqref{EQ} is a model independent equation although its {\it solution} for $\phi_k$ requires specifying $N_{{ep}}$ and a model of inflation; $H_k$ and $N_{ke}$ are model dependent quantities. Thus, after finding $\phi_k$,  we can proceed to determine all inflationary parameters and observables.
 
\noindent
To find the value of $a_p$  we solve the Friedmann equation which can be written in the form
\begin{equation}
k_p=H_0\sqrt{\frac{\Omega_{md,0}}{a_p}+\frac{\Omega_{rd,0}}{a_p^2}+\Omega_{de}a_p^2}\,\approx H_0\sqrt{\frac{\Omega_{md,0}}{a_p}+\frac{\Omega_{rd,0}}{a_p^2}}\;, 
 \label{Frid}
\end{equation}
where $k_p=0.05/Mpc \approx 1.3105 \times 10^{-58}$ (see Table \ref{tableparameters} to find the numerical values of the other parameters used in our calculations). The solution of Eq.~\eqref{Frid} for $a_p$ is $a_p=3.6512\times 10^{-5}$ from where we get $N_{p0}=10.2$ for the number of e-folds from $a_p$ to $a_0$.
Note also that Eq.~\eqref{EQ} incorporates knowledge from the present universe, in the determination of $a_p$, of the early universe, when considering the scale $k$ during inflation, and also of the CMB epoch by the presence of the scalar power spectrum amplitude $A_s$ through $H_k$.

From Eq.~\eqref{EQ} and $H_k=\sqrt{8\pi^2 \epsilon_k A_s}$ we can get an expression for $N_{ke}+N_{ep}$ in terms of the tensor-to-scalar index $r\equiv 16\epsilon_k$
\begin{equation}
N_{ke}+N_{ep}=\frac{1}{2}\ln\left(\frac{\pi^2 a_p^2 A_s}{2 k_p^2}r\right)\;.
 \label{erre}
\end{equation}
Imposing a bound $b$ to $r$ we get a general bound for $N_{ke}+N_{ep}$
\begin{equation}
r< b\,\,\, \Rightarrow \,\,\, N_{ke}+N_{ep}<\frac{1}{2}\ln\left( \frac{\pi^2 a_p^2 A_s}{2 k_p^2}\,b\right)\;, 
 \label{generalboundNk}
\end{equation}
for the particular value $b=0.063$     \cite{Aghanim:2018eyx}, \cite{Akrami:2018odb} we get the present bound for $N_{ke}+N_{ep}$
\begin{equation}
r< 0.063 \quad \Rightarrow \quad N_{ke}+N_{ep}<112.5,\quad at\quad k=k_p\;.
 \label{particularboundNk}
\end{equation}
This is a model independent result, it follows from Eq.~\eqref{EQ}, phenomenological parameters and the bound for $r$ without specifying any model of inflation. We rely that Eq.~\eqref{Frid} describes well the Universe when the scale $k_p$ re-renters the horizon. Eq.~\eqref{Frid} depends on post-inflationary physics through $\Omega_{md,0}$, $\Omega_{rd,0}$ and $H_0$ however, the bound given by Eq.~\eqref{particularboundNk} does not depend on the nature of reheating or on a specific model of inflation. Also, when using the expression $H_k=\sqrt{8\pi^2 \epsilon_k A_s}$ above we have in mind single-field models of inflation, it would be interesting to see how the results presented here are modified for non canonical models of inflation or multifield inflation.

\begin{figure}[tb]
\captionsetup{justification=raggedright,singlelinecheck=false}
\par
\includegraphics[width=12cm]{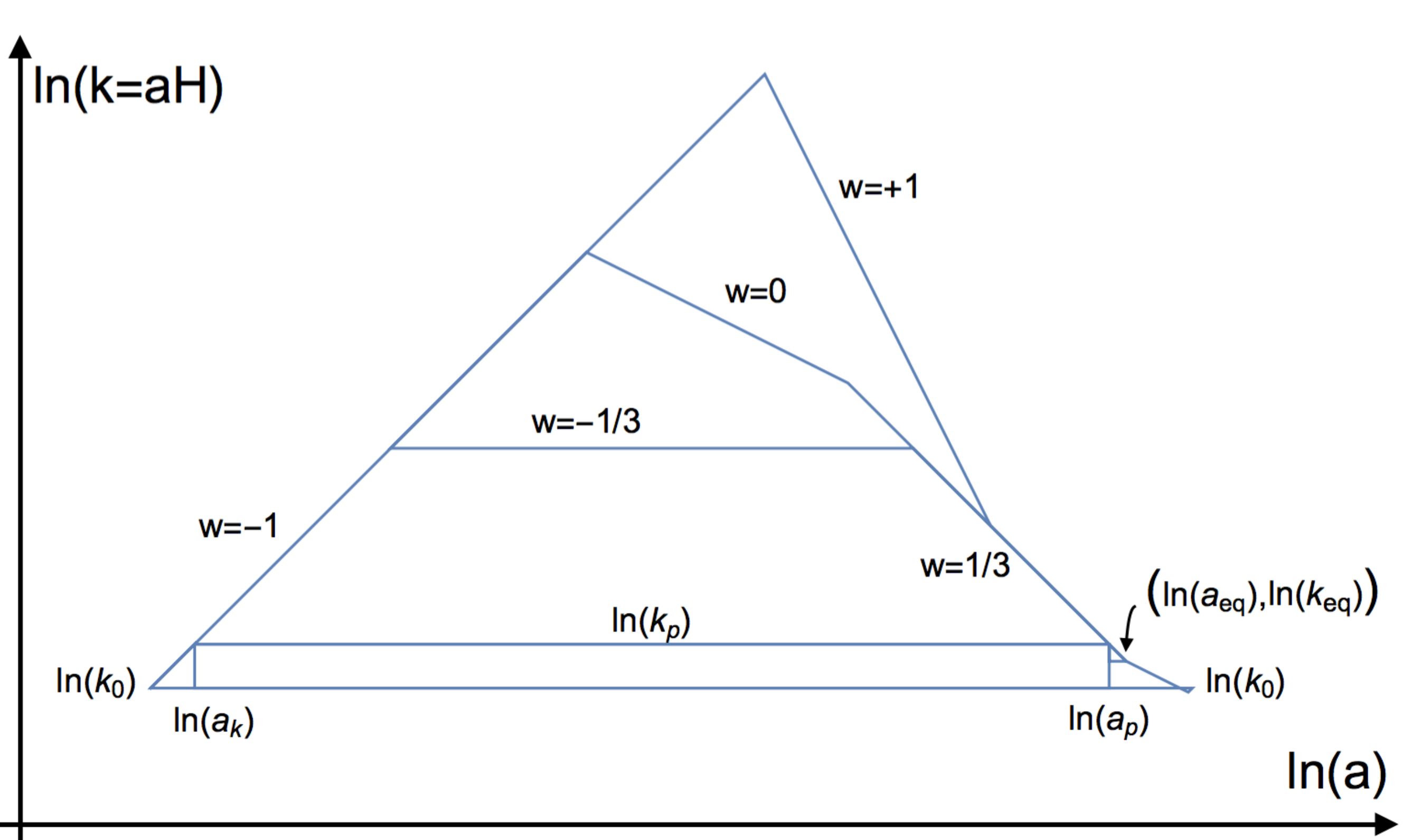}
\caption{\small Diagram for the evolution of the comoving scale of wavenumber $k\equiv a H_k$ showing $\ln k$ as a function of the logarithm of the scale factor $\ln a$ for three possible examples of reheating: those described by an EoS $\omega_{re}$ equal to -1/3, 0 and 1. The diagram is fixed by the radiation line of $\omega_{rad}=1/3$ (fixed by the pivot point at $(\ln a_p,\ln k_p)$) an the inflationary line of $\omega_{inf}=-1$ fixed by the length of the horizontal  line $\ln (k_p)$. All other lines are drawn in reference to this fixed framework   \cite{German:2020kdp}, \cite{German:2020dih}.}
\label{diagrama}
\end{figure}

We can also calculate a model independent bound to the size of the patch of the universe from which our present observable universe originates. We adapt Eq.~\eqref{EQ} to this situation:
\begin{equation}
\ln\left(\frac{a_{0}}{a_{k}}\right)=N_{ke}+N_{e0}\,,\quad at\quad k=k_0\;,
\label{eq0}
\end{equation}
where now $k=k_0$ is the scale at horizon crossing during inflation which gave rise to our observable universe ($k_0\equiv a_0\,H_0$ is the present scale wavenumber)  $a_0$ is, as usual, the present scale factor $a_0=1$, and $N_{ke}+N_{e0}$ is the number of e-folds from $a_k$ up to the end of inflation {\it plus} the number of e-folds from the end of inflation to the present. From Eq.~\eqref{EQ} and from the bound for $N_{ke}+N_{ep}$ follows that at the scale $k=k_0$
\begin{equation}
a_k = a_0\, e^{-(N_{ke}+N_{k_0})} > a_0\, e^{-128.1}\approx 2.3 \times 10^{-56}a_0\;.
 \label{ak}
\end{equation}
\\
Note that we have added 5.4 e-folds to the upper bound of 112.5  because there are 5.4 e-folds coming from the time when observable scales the size of the present scale left the horizon at $a_{k_0}$ to the time when scales the size of the pivot scale left the horizon at $a_{k_p}$ during inflation  (l.h.s. corner of Fig~\ref{diagrama}) and $N_{p0}\equiv \ln\frac{a_0}{a_p}= 10.2$ e-folds from the pivot scale up to the present scale with wavenumber mode $k_0$ (r.h.s. corner of Fig~\ref{diagrama}). 
Thus, the total number of e-folds which our observable universe has expanded since the beginning of observable inflation to the present is bounded as
\begin{equation}
N_{total} \leq 128.1\;.
 \label{Nt}
\end{equation}
This is a general result which any model of inflation should satisfy. This result can give a model independent lower bound to the size of the universe at the beginning of observable inflation.
If the diameter of the observable universe is $8.8 \times 10^{26}m$ then at the scale $k$ the size of the universe from which ours originates was bigger than $ 2.05 \times 10^{-29}m$. Thus, at the scale $k=k_0$ the universe diameter was 
{\it at least} $1.27 \times 10^{6}$ times bigger than the Planck length. 
\begin{table*}[htbp!]
\captionsetup{justification=raggedright,singlelinecheck=false}
\par
\begin{tabular}{ccc}
    $Parameter$ & \quad Usually given as   &\,\, $Dimensionless$\\ \hline
   & \quad   & \quad  \\[-2mm]
  $H_0$ & \quad $100\,h\frac{km}{s}/Mpc$  & \quad $8.7426 \times 10^{-61}\,h$ \\[2mm]
  $T_0$ & \quad $2.725\, K$  & \quad $9.6235\times 10^{-32}$\\[2mm]
    $A_s$ & \quad $2.0968 \times 10^{-9}$  & \quad $2.0991 \times 10^{-9}$\\[2mm]
   $k_p$ & \quad $0.05/Mpc$  & \quad $1.3105\times 10^{-58}$\\[2mm]
$a_p$ & \quad $-$  & \quad $3.6512\times 10^{-5}$\\[2mm]
$\Omega_{md,0}$  & \quad $$0.315$$ & \quad $0.315$\\[2mm]
$\Omega_{rd,0}$   & \quad $5.443\times 10^{-5}$  & \quad $5.443\times 10^{-5}$\\[2mm]
$\Omega_{de}$  & \quad $0.685$  & \quad $0.685$\\[2mm]
$g_{s,re}=g_{re}$  & \quad $106.75$  & \quad $106.75$\\[2mm]
 \end{tabular}
 \caption{\label{tableparameters} For easy reference this table collects numerical values of parameters used in the paper. Dimensionless quantities have been obtained by working in Planck mass units, where $M_{pl}=2.4357\times 10^{18} GeV$ and set $M_{pl}=1,$ the pivot scale $k_p\equiv a_p H_p=0.05\frac{1}{Mpc}$, used in particular by the Planck collaboration, becomes a dimensionless number given by $k_p\approx 1.3105\times 10^{-58}$. To calculate $a_p$ we have to specify $h$ for the Hubble parameter $H_0$ at the present time. We  take the value given  by Planck $h=0.674$ for definitiveness.  The solution of Eq.~\eqref{Frid} for $a_p$ is $a_p=3.6512\times 10^{-5}$ from where we get $N_{p0}=10.2$ for the number of e-folds from $a_p$ to $a_0$.\\ }
\end{table*}
\section {\bf Formulas for the reheating and radiation epochs}\label{RR}
Here we give formulas for the number of e-folds during reheating $N_{re}$ and also for the number of e-folds during the radiation dominated epoch $N_{rd}$. The standard way to proceed is to solve the fluid equation with the assumption of a constant equation of state parameter $\omega$, this gives the number of e-folds during reheating in terms of the energy densities as follows
\begin{equation}
N_{re}\equiv\ln\frac{a_{re}}{a_{e}} = [3(1+\omega_{re})]^{-1}\ln[\frac{\rho_e}{\rho_{re}}]\;,
\label{NRE1}
\end{equation}
\\
where $\rho_{e}$ is the energy density at the end of inflation and $\rho_{re}$ the energy density at the end of reheating
\begin{equation}
\rho_{re} = \frac{\pi^2 g_{re}}{30} T_{re}^4\;,
\label{RO1}
\end{equation}
with $g_{re}$ the number of degrees of freedom of species at the end of reheating.  To proceed we assume entropy conservation after reheating, this assumption establish another expression involving $T_{re}$ which can be substituted in Eq.~\eqref{RO1} and then in Eq.~\eqref{NRE1}
\begin{equation}
g_{s,re}T_{re}^3=\left(\frac{a_0}{a_{eq}}\right)^3\left(\frac{a_{eq}}{a_{re}}\right)^3\left(2 T_0^3+6\times \frac{7}{8}T_{\nu,0}^3\right)\;,
\label{entropy}
\end{equation}
where $g_{s,re}$ is the entropy number of degrees of freedom of species after reheating, $T_0=2.725 K$ and the neutrino temperature is $T_{\nu,0}=(4/11)^{1/3} T_0$. The number of e-folds during radiation domination $N_{rd}\equiv\ln\frac{a_{eq}}{a_{re}}$ follows from Eqs.~\eqref{NRE1} and \eqref{entropy}
\beq
\label{NRD}
N_{rd}= -\frac{3(1+\omega)}{4}N_{re}+\frac{1}{4} \ln[\frac{30}{g_{re} \pi^2}] +\frac{1}{3} \ln[\frac{11 g_{sre}}{43}]+\ln[\frac{a_{eq}\, \rho_e^{1/4}}{a_0\,T_0}]\;.
\eeq
We can finally obtain an expression for the number of e-folds during reheating $N_{re}$ by combining Eqs.~\eqref{EQ} and \eqref{NRD}, the result is \cite{German:2020dih}
\beq
\label{NRE}
N_{re}= \frac{4}{1-3\, \omega}\left(-N_{ke}-\frac{1}{3} \ln[\frac{11 g_{s,re}}{43}]-\frac{1}{4} \ln[\frac{30}{\pi^2 g_{re} } ] -\ln[\frac{\rho^{1/4}_e k}{H_k\, a_0 T_0} ]\right)\;.
\eeq
A final quantity of physical relevance is the thermalization temperature at the end of the reheating phase
\beq
\label{TRE}
T_{re}=\left( \frac{30\, \rho_e}{\pi^2 g_{re}} \right)^{1/4}\, e^{-\frac{3}{4}(1+\omega_{re})N_{re}}\,.
\eeq
\noindent 
This is a function of the number of $e$-folds during reheating. It can also be written as an equation for the parameter $\omega_{re}$
\begin{equation}
\omega_{re}=\frac{1}{3}+\frac{4 \bar{N}_{re}}{3\left(-\bar{N}_{re}+\frac{1}{4}\ln[\frac{\pi^2\,g_{re}} {30\rho_e}\, {T_{re}}^4]\right)}\;, 
 \label{w}
\end{equation}
where $\bar{N}_{re}$ is just the term in the brackets of Eq.~\eqref{NRE} and is independent of $\omega_{re}$.
From Eq.~\eqref{w} we can rewrite the equations for $N_{re}$ and $N_{rd}$ as functions of  $T_{re}$ and $n_s$  and of $T_{re}$, respectively
\begin{equation}
N_{re}= \bar{N}_{re}-\frac{1}{4}\ln[\frac{\pi^2\,g_{re}} {30\rho_e}] - \ln[T_{re}]\;, 
 \label{NRET}
\end{equation}
\begin{equation}
N_{rd}= \ln[\frac{a_{eq}}{a_0\,T_0}]+\frac{1}{3} \ln[\frac{11 g_{s,re}}{43}]+\ln[T_{re}]\;.
 \label{NRDT}
\end{equation}
The dependence on $n_s$ occurs because $n_s$ is related to $\phi_k$ through the expression for the spectral index and $\phi_k$ is present in the terms $N_{ke}$ and $H_{k}$ contained in the definition of $\bar{N}_{re}$ above.
From these two equations we see that $N_{re}+N_{rd}$ is independent of $T_{re}$, equivalently $\omega_{re}$ independent. Thus, the sum $N_{re}+N_{rd}$ only depends on $\phi_k$, the value of the inflaton at $k$ (equivalently on $n_s$) and also of parameters present in the potential defining the model, if any.
\section {\bf Mutated Hilltop Inflation type models}\label{MUT}
\begin{figure}[tb]
\captionsetup{justification=raggedright,singlelinecheck=false}
\par
\includegraphics[width=12cm]{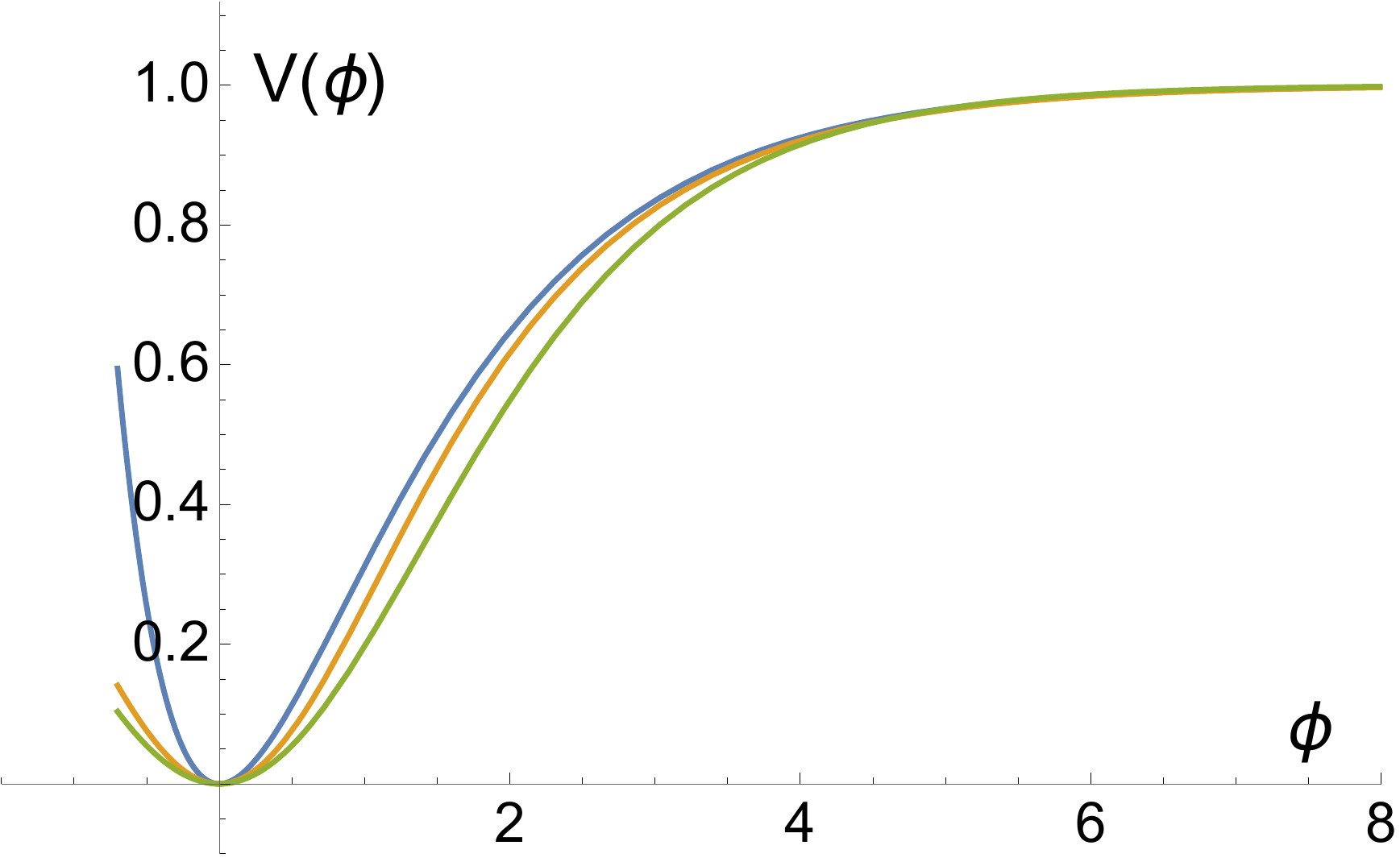}
\caption{\small Schematic plot of (from top to bottom) the Satorobinsky, PPB and AFMT potentials given by Eqs.~(\ref{staropot}), (\ref{ppb}) and (\ref{afmt}) respectively as functions of $\phi$ for an inflaton field rolling from the right. These model are degenerated at horizon crossing at $\phi_k$ during the inflationary epoch but can be distinguished during reheating (see Table~\ref{table4}).}
\label{pots}
\end{figure}

\begin{figure}[tb]
\captionsetup{justification=raggedright,singlelinecheck=false}
\par
\includegraphics[width=12cm]{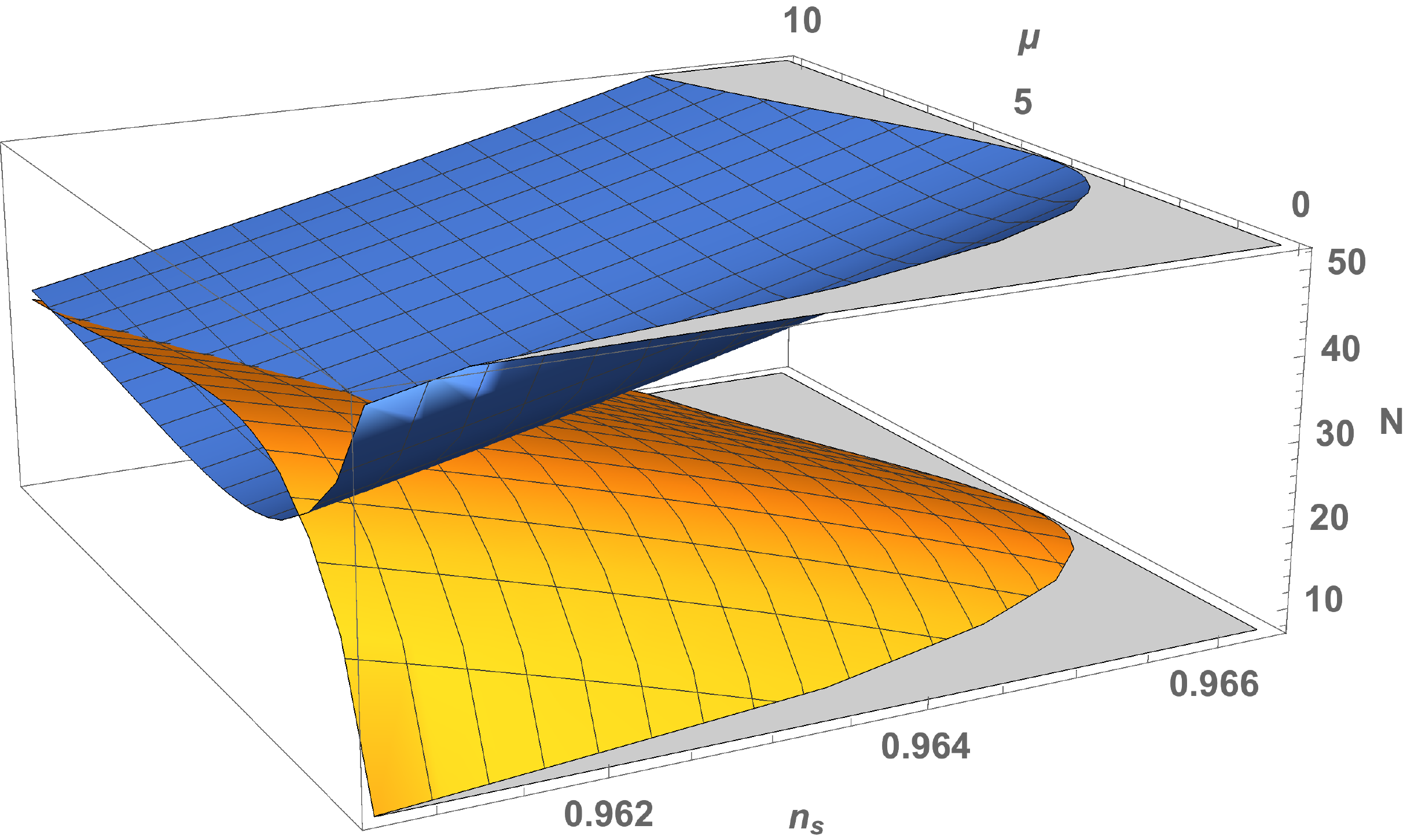}
\caption{\small Plot of the number of e-folds during reheating $N_{re}$ (bottom surface) and during radiation domination $N_{rd}$ as  functions of the spectral index $n_s$ and of the mass parameter $\mu$ for the PPB model given by the potential of Eq.~(\ref{ppb}). Quantities dobtained from the bounds for $n_s$ and $\mu$ are given in Tables~\ref{table2} and \ref{table3}.}
\label{NreNrd}
\end{figure}

As the measurements of cosmological observables become more and more accurate, the number of models capable of describing them is reduced. However, a certain degeneration of models persists and it seems impossible to break it using observations from the inflationary stage only. Next we show with several examples how a knowledge of the reheating epoch allows to break the degeneration and distinguish between models with very similar predictions for the inflationary era. Here, we apply the results discussed in the previous sections to Mutated Hilltop Inflation type models starting with the Pal, Pal, Basu (PPB) model.
\\
{\it The PPB model.-}
The PPB model is given by the potential \cite{Pal:2009sd}, \cite{Pal:2010eb}
\begin{equation}
\label{ppb}
V= V_0 \left(1- \sech\left(\frac{\phi}{\mu}\right)\right),
\end{equation}
and shown in Fig.~\ref{pots}.
The number of e-folds during inflation $N_{ke}$ can be calculated in closed form with the result
\beq
\label{Nkeppb}
N_{ke} = 2\mu^2 \ln\left(\frac{\cosh\left(\frac{\phi_e}{2\mu}\right)}{\cosh\left(\frac{\phi_k}{2\mu}\right)}\right) + \cosh\left(\frac{\phi_k}{\mu}\right)  - \cosh\left(\frac{\phi_e}{\mu}\right).
\eeq
The field at the end of inflation $\phi_e$ is given by the solution to the condition $\epsilon =1$. The solution is very involved and is given by
\beq
\resizebox{1.00\textwidth}{!}{$
\label{fieppb}
\cosh\left(\frac{\phi_e}{\mu}\right) = \frac{-36\mu^4(3+2\mu^2)^2}{
12\mu^4(3+2\mu^2)^2-4\mu^4(99+72\mu^2+4\mu^4)R^{1/3}+2\mu^2(-9+60\mu^2+4\mu^4)R^{2/3}+(3+2\mu^2)R^{4/3}-R^{5/3}}\,,$}
\eeq
where $R=2\mu^3\left(4\mu(9+\mu^2)+3\sqrt{6}\sqrt{-1+22\mu^2+4\mu^4}\right)$.
We cannot solve in general Eq.~\eqref{EQ} for $\phi_k$ and arbitrary $\mu$ but from the expression for the spectral index $n_s=1+2\eta-6\epsilon$ we can write $\phi_k$ in terms of $n_s$ and use bounds on $n_s$ to study the model. Thus,
\beq
\resizebox{0.95\textwidth}{!}{$
\label{fikppb}
\cosh\left(\frac{\phi_k}{\mu}\right)=\frac{2+\bar \mu^2+\left(8+\bar \mu^2(39+\bar \mu^2(6+\bar \mu^2))+  i\, 3\sqrt{3}\bar \mu^2\sqrt{17+\bar \mu^2(75+\bar \mu^2(15+\bar \mu^2))}\right)^{1/3} + c.c.}{3\bar \mu^2}\,,$}
\eeq
where $\bar \mu^2\equiv \mu^2(1-n_s)$. The PPS model is very well approximated near the origin by a quadratic potential thus, it makes sense to study the reheating epoch with and EoS given by $\omega_{re}=0$ \cite{Turner:1983he}. 
In Fig.~\ref{NreNrd} we plot the number of e-folds during reheating, Eq.~\eqref{NRE}, and during radiation domination, Eq.~\eqref{NRD}, as functions of the mass parameter $\mu$ and the spectral index $n_s$. From the Planck bounds for the spectral index \cite{Aghanim:2018eyx}, \cite{Akrami:2018odb} $0.9607<n_s<0.9691$ and from the Fig.~\ref{NreNrd} we see that the condition $N_{re}\geq 7$ implies $0.9607<n_s<0.9664$ and $5\times 10^{-3}<\mu<10$. The lower bound $N_{re}\geq 7$ comes from a recent lattice simulation for a quadratic monomial and potentials flattening at large field values like the PPB  \cite{Antusch:2020iyq}. The bound $N_{re}\geq 7$ is very conservative with the expectation that it should be much larger, however, the numerical results where unable to reach the radiation dominated era for this case.
In Tables \ref{table2} and \ref{table3} we give bounds for quantities of interest during inflation, reheating and radiation for various values of $\mu$. 
 \begin{center}
 \captionsetup{justification=raggedright,singlelinecheck=false}
\par
\addtolength{\tabcolsep}{-7pt}
\begin{table*}[htbp!]
\begin{tabular}{ccccccc}
    $\mu$ & \quad $n_s$  &\,\, $r$ & \,\,$\alpha$ & \,\,$N_{ke}$ & \,\,$N_{re}$ & \,\, $N_{rd}$\\ \hline\\[0.1mm]
$0.5$ & \quad $(0.9607,0.9631)$  & \quad $(7.6,6.7)\times 10^{-4}$ & \quad $(-7.8,-6.8)\times 10^{-4}$ & \quad $(49.9,53.2)$ & \quad $(20.5,7)$ & \quad $(42.0,52.1)$\\[2mm]
$1$ & \quad $(0.9607,0.9642)$  & \quad $(2.9,2.4)\times 10^{-3}$ & \quad $(-7.8,-6.5)\times 10^{-4}$ & \quad $(48.7,53.7)$ & \quad $(27.1,7)$ & \quad $(37.3,52.3)$\\[2mm]
$4$ & \quad $(0.9607,0.9664)$  & \quad $(2.9,2.2)\times 10^{-2}$ & \quad $(-8.4,-6.1)\times 10^{-4}$ & \quad $(46.8,54.7)$ & \quad $(38.8,7)$ & \quad $(28.7,52.4)$\\[2mm]
$7$ & \quad $(0.9607,0.9659)$  & \quad $(5.6,4.6)\times 10^{-2}$ & \quad $(-8.4,-6.3)\times 10^{-4}$ & \quad $(47.9,55.0)$ & \quad $(38.8,7)$ & \quad $(31.1,52.4)$\\[2mm]
 \end{tabular}
 \caption{\label{table2} For a model similar to PPB where the potential can be approximated by a quadratic monomial at the origin recent lattice simulations  \cite{Antusch:2020iyq} suggest that 
there are {\it at least} 7 e-folds previous to entering the radiation era (see upper left hand panel of Fig.~1 of \cite{Antusch:2020iyq}). This is a very conservative lower bound and can be much larger that 7. The lower bound for $n_s$ comes from the Planck collaboration \cite{Aghanim:2018eyx}, \cite{Akrami:2018odb} while the upper bound as well as the bounds for the mass parameter $\mu$ are obtained imposing the condition $N_{re}>0$ in Eq.~\eqref{NRE} and
can be read directly from Fig.~\ref{NreNrd}. Bounds are given for the tensor-to-scalar ratio $r$, running $\alpha$, number of e-folds during inflation $N_{ke}$, during reheating $N_{re}$ and during the radiation dominated $N_{rd}$ epochs.
\\}

\end{table*}
\end{center}
 \begin{center}
 \captionsetup{justification=raggedright,singlelinecheck=false}
\par
\addtolength{\tabcolsep}{-7pt}
\begin{table*}[htbp!]
\begin{tabular}{ccccc}
    $\mu$ & \quad $n_s$  &\,\, $V_k^{1/4}\,(GeV)$ & \,\,$H_k\, (GeV)$ & \,\,$T_{re}\,(GeV)$\\ \hline\\[0.1mm]
$0.5$ & \quad $(0.9607,0.9631)$  & \quad $(5.4,5.2)\times 10^{15}$ & \quad $(6.8,6.4)\times 10^{12}$ & \quad $(4.5\times 10^{8},1.1\times 10^{13})$\\[2mm]
$1$ & \quad $(0.9607,0.9641)$   & \quad $(7.5,7.2)\times 10^{15}$ & \quad $(1.3,1.2)\times 10^{13}$ & \quad $(4.1\times 10^{6},1.3\times 10^{13})$\\[2mm]
$4$ & \quad $(0.9607,0.9664)$   & \quad $(1.3,1.3)\times 10^{16}$ & \quad $(4.2,3.7)\times 10^{13}$ & \quad $(7.3\times 10^{2},1.5\times 10^{13})$\\[2mm]
$7$ & \quad $(0.9607,0.9659)$  & \quad $(1.6,1.5)\times 10^{16}$ & \quad $(5.9,5.2)\times 10^{13}$ & \quad $(8.1\times 10^{3},1.5\times 10^{13})$\\[2mm]
 \end{tabular}
 \caption{\label{table3} This table is a continuation of Table~\ref{table2} for the PPB model given by Eq.~(\ref{ppb}). Bounds are given for the scale of inflation $V_k^{1/4}$, Hubble function $H_k$ and reheat temperature $T_{re}$.
\\}
\end{table*}
\end{center}

{\it The AFMT model.-}
Here, we apply the results discussed in the previous sections to the AFMT model given by the potential \cite{Antusch:2020iyq}
\beq
\label{afmt}
V(\phi,X)= \frac{1}{p}\Lambda^4 \tanh^p\left(\frac{\left |\phi\right |}{M}\right)+\frac{1}{2}g^2\phi^2X^2,
\eeq
where $M$ and $\Lambda$ are mass scales and $g$ is a dimensionless coupling parameter. The first term is the inflationary potential and the second gives the interaction of the inflaton with a light field $X$ to which energy is transfered.
The number of e-folds during inflation $N_{ke}$ can be calculated in closed form with the result
\beq
\label{NkeAFMT}
N_{ke}= -\int_{\phi_k}^{\phi_e}\frac{V}{V'}d\phi = \frac{M^2}{4p}\left(\cosh\left(\frac{2\phi_k}{M}\right)-\cosh\left(\frac{2\phi_e}{M}\right)\right).
\eeq
The field at the end of inflation $\phi_e$ is given by the solution to the condition $\epsilon =1$
\beq
\label{fieAFMT}
\sinh\left(\frac{2\phi_e}{M}\right) = \frac{\sqrt{2}\,p}{M}\,,
\eeq
From the expression for the spectral index  we obtain $\phi_k$ in terms of $n_s$ and use bounds on $n_s$ to study the model thus,
\beq
\label{fikAFMT}
\phi_k=\frac{M}{2}\ln\left(\frac{4p+\left(\bar M^4+4p^2(4+\bar M^2)\right)^{1/2}+2 \sqrt{p}\left(8p+\bar M^2 p+2\left(\bar M^4+4p^2(4+\bar M^2)\right)^{1/2}\right)^{1/2}}{\bar M^2}\right)\,,
\eeq
where $\bar M^2\equiv M^2(1-n_s)$. 
We can also have another expression for $\phi_k$ by solving in terms of the number of e-folds during inflation $N_{ke}$, from Eq.~\eqref{NkeAFMT}
\beq
\label{fikNke}
\cosh\left(\frac{2\phi_k}{M}\right)=\left(\frac{4p}{M^2}N_{ke}+\cosh\left(\frac{2\phi_e}{M}\right)\right)\,,
\eeq
Full equations for $n_s$ and $r$ in terms of $N_{ke}$ can be written with the following large-$N_{ke}$ expansions
\beq
\label{nsN}
n_s\approx1-\frac{2}{N_{ke}}\,,
\eeq
\beq
\label{rN}
r\approx \frac{2M^2}{N^2_{ke}}\,.
\eeq
{\it The Starobinsky model.-}
 The potential of the Starobinsky model \cite{Starobinsky:1980te,Mukhanov:1981xt,Starobinsky:1983zz} is given by \cite{Whitt:1984pd}:
\beq
\label{staropot}
V= V_0 \left(1- e^{-\sqrt{\frac{2}{3}}\phi} \right)^2,
\eeq
with Hubble function
\beq
\label{Hk}
H_k=\sqrt{8\pi^2 \epsilon_k A_s}=\sqrt{\frac{32 A_s}{3}}\frac{\pi}{e^{\sqrt{\frac{2}{3}}\phi_k}-1}\,,
\eeq
where $\epsilon_k$ is the slow-roll parameter $\epsilon_1\equiv\frac{1}{2}\left(\frac{V'}{V}\right)^2$ at $\phi=\phi_k$. The number of e-folds $N_k$ follows easily 
\beq
\label{Nk}
N_k = -\int_{\phi_k}^{\phi_e}\frac{V}{V'}d\phi = \frac {1}{4} \left( 3e^{\sqrt{\frac{2}{3}}\phi_k} -\sqrt{6} \,\phi_k \right)-\frac {1}{4} \left( 3e^{\sqrt{\frac{2}{3}}\phi_e} -\sqrt{6} \,\phi_e \right),
\eeq
where $\phi_e$ signals the end of inflation. It is given by the solution to the equation $\epsilon\equiv \frac{1}{2}\left(\frac{V_{\phi}}{V}\right)^2=1$: $\phi_e= \sqrt{\frac {3}{2}}\ln\left(1+\frac{2}{\sqrt{3}}\right)$.
Notice that in the Starobinsky model there are no further parameters apart from the overall scale $V_0$ which is fixed by the scalar amplitude.
 \begin{center}
 \captionsetup{justification=raggedright,singlelinecheck=false}
\par
\addtolength{\tabcolsep}{-7pt}
\begin{table*}[htbp!]
\begin{tabular}{ccccccccc}
    $Model$ & \quad $n_s$  &\,\, $r$ & \,\,$\alpha$ & \,\,$N_{ke}$ & \,\,$N_{re}$ & \,\, $N_{rd}$ &\,\, $V_k^{1/4}\,(GeV)$ & \quad $T_{re}$(GeV)\\ \hline\\[0.1mm]
  Starobinsky & \quad $0.9649$  & \quad $0.00351$ & \quad $-6.2 \times 10^{-4}$ & \quad $53.7$ & \quad $6.7$ & \quad $52.7$ & \quad $7.9 \times 10^{15}$ & \quad $2.0 \times 10^{13}$\\[2mm]
PPS, $\mu=1.2395$ & \quad $0.9649$  & \quad $0.00351$ & \quad $-6.3 \times 10^{-4}$ & \quad $54.2$ & \quad $5.6$ & \quad $53.4$ & \quad $7.9 \times 10^{15}$ & \quad $4.0 \times 10^{13}$\\[2mm]
PPS, $\mu=1.3786$ & \quad $0.9649$  & \quad $0.00428$ & \quad $-6.3 \times 10^{-4}$ & \quad $53.9$ & \quad $7$ & \quad $52.3$ & \quad $8.3 \times 10^{15}$ & \quad $1.4 \times 10^{13}$\\[2mm]
AFMT, $M=2.4187$ & \quad $0.9649$  & \quad $0.00351$ & \quad $-6.2 \times 10^{-4}$ & \quad $56.6$ & \quad $-3.8$ & \quad $60.3$ & \quad $7.9 \times 10^{15}$ & \quad $4.1 \times 10^{16}$\\[2mm]
 \end{tabular}
 \caption{\label{table4} In rows 1, 2 and 4 we compare the Satorobinsky, PPB and AFMT potentials given by Eqs.~(\ref{staropot}), (\ref{ppb}) and (\ref{afmt}) respectively (for the arbitrarily chosen central value of the spectral index given by Planck \cite{Aghanim:2018eyx}, \cite{Akrami:2018odb}) by fixing the mass scales $\mu$ and $M$ for the PPB and AFMT models in such a way that the tensor-to-scalar ratio $r$ gets the same value for all three models. We see that quantities at $\phi_k$ during inflation (the running $\alpha$ and the scale of inflation $V_k^{1/4}$)
are essentially the same but differences arise during the reheating epoch. 
Fixing $\mu$ in row 3  in such a way that the minimum $N_{re}\geq7$ is obtained changes completely the prediction for $r$ while in row 4 fixing $M$ to get the same $r$ as in rows 1 and 2 gives a negative (unacceptable) $N_{re}$. The conclusion is that the degeneracy present in models of inflation cannot be resolved by considering the inflationary epoch itself but requires knowledge of the reheating epoch.
\\}
\end{table*}
\end{center}
In Table \ref{table4} we compare the a Starobinsky, PPS and ATMF models of inflation for the (arbitrarily chosen) central value $n_s=0.9649$ and for values of the mass parameters $\mu$ and $M$ such that the tensor-to-scalar index $r$ has the same value for all the models; we see that it would be very difficult to distinguish between these models by looking at the inflationary observables only. It becomes clear how the knowledge of the reheating epoch is essential to break the degeneracy among these models (see Table~\ref{table4} caption).
\section {\bf Conclusions}\label{Con}
We have studied model independent results for the inflationary epoch following from the formula given by Eq.~\eqref{EQ}. We have in particular established an equation (Eq.~\eqref{erre}) for the the number of e-folds $N_{ke}+N_{ep}$, from $a_k$ during inflation to the pivot scale at $a_p$ in terms of the tensor-to-scalar ratio $r$. From a bound $b$ for $r$ follows a general bound for $N_{ke}+N_{ep}$ (Eq.~\eqref{generalboundNk})  which at present is $r<0.063$ implying $N_{ke}+N_{ep} < 112.5$ at the scale $k=k_p$ or $N_{ke}+N_{e0} < 128.1$ at the present scale $k_0$. These are all model independent results in the sense that no model of inflation has been used to obtain them. At the end of Section \ref{RI} we also give a model independent lower bound to the size of patch of the universe from where our observable universe comes from. We have also discussed the degeneracy of models of inflation arguing that it is not possible to break their degeneracy by looking at the inflationary epoch only. We study three simple models giving essentially the same observables during inflation and discussing how the knowledge of the reheating epoch is necessary to break their degeneracy. These results are summarize in Table \ref{table4}.

\acknowledgments

We acknowledge financial support from UNAM-PAPIIT,  IN104119, {\it Estudios en gravitaci\'on y cosmolog\'ia}.

\end{document}